\documentclass[aps,prl,reprint,10pt]{revtex4-1}
\pdfoutput=1
\usepackage[breaklinks]{hyperref}
\usepackage[pdftex]{graphicx}
\usepackage[applemac]{inputenc}
\usepackage{natbib}
\usepackage{bm}
\usepackage{amssymb}
\usepackage{amsbsy}
\usepackage[usenames,dvipsnames]{xcolor}
\usepackage{mathrsfs,mathcomp}
\usepackage[normalem]{ulem}
\usepackage{amsmath}
\usepackage{amsfonts}
\usepackage{mathrsfs}
\usepackage{subfigure}
\usepackage{ifsym} 
\usepackage{multirow}
\usepackage{textgreek}
\usepackage{textcomp}
\usepackage{gensymb}

\newcommand{\SIC}{^\circ\!\textrm{C}}

\newcommand{\SImuL}{\textrm{\textmu{}L}}

\newcommand{\SIkg}{\textrm{kg}}

\newcommand{\SIm}{\textrm{m}}

\newcommand{\SImum}{\textrm{\textmu{}m}}

\begin{document}

\title{Surfactant-driven flow transitions in evaporating droplets}

\author{Alvaro Marin, Robert Liepelt, Massimiliano Rossi, Christian J. K\"ahler} 

\affiliation{Institute for Fluid Mechanics and Aerodynamics, Bundeswehr University Munich, Germany}

\begin{abstract}
An evaporating droplet is a dynamic system in which flow is spontaneously generated to minimize the surface energy, dragging particles to the borders and ultimately resulting in the so-called ``coffee-stain effect''. The situation becomes more complex at the droplet's surface, where surface tension gradients of different nature can compete with each other yielding different scenarios. With careful experiments and with the aid of 3D particle tracking techniques, we are able to show that different types of surfactants turn the droplet's surface either rigid or elastic, which alters the evaporating fluid flow, either enhancing the classical coffee-stain effect or leading to a total flow inversion. Our measurements lead to unprecedented and detailed measurements of the surface tension difference along an evaporating droplet's surface with good temporal and spatial resolution. 

\end{abstract}
\maketitle
\paragraph{\bf Introduction.- }
Evaporating capillary droplets might appear simple systems but they hide surprisingly complex phenomena. One of the most fascinating effects are the spontaneous evaporation-driven flows that can be generated inside the droplet. The most dominant of them is the one giving rise to the so-called ``coffee-stain effect'':\cite{Deegan:1997vb,deegan2000contact,deegan2000pattern} A capillary flow refills the corners of the droplet, dragging any dispersed particle in the liquid towards the contact line, where they get trapped. The outcome is a characteristic ring-shaped stain which we can see often on tables where a spilled drop of coffee has evaporated. 
There are however different kind of spontaneous flows that can be induced within an evaporating drop. Temperature differences might eventually develop along the evaporating droplet surface, leading to surface tension differences and therefore to a Marangoni flow. \cite{davis1987thermocapillary} Modeling the temperature and the flow field, Hu and Larson \cite{Hu:2005dv} predicted that a significant Marangoni flow should develop at a water drop's surface evaporating on a glass substrate. Since such Marangoni stresses induce a surface flow in the opposite direction of the bulk's capillary flow, many authors have claimed that it could eventually be used to reverse the coffee-stain effect. \cite{Hu:2006tx,ristenpart2007influence} Nonetheless, few have managed to visualize such a process, \cite{Kajiya:2008uk} and even less to quantify it. Consequently, most of the studies in the literature rely only on the observation of the final stain to infer on the complex phenomena occurring during the droplet evaporation. 

\begin{figure}[h]
\centering
\includegraphics[width=.4\textwidth]{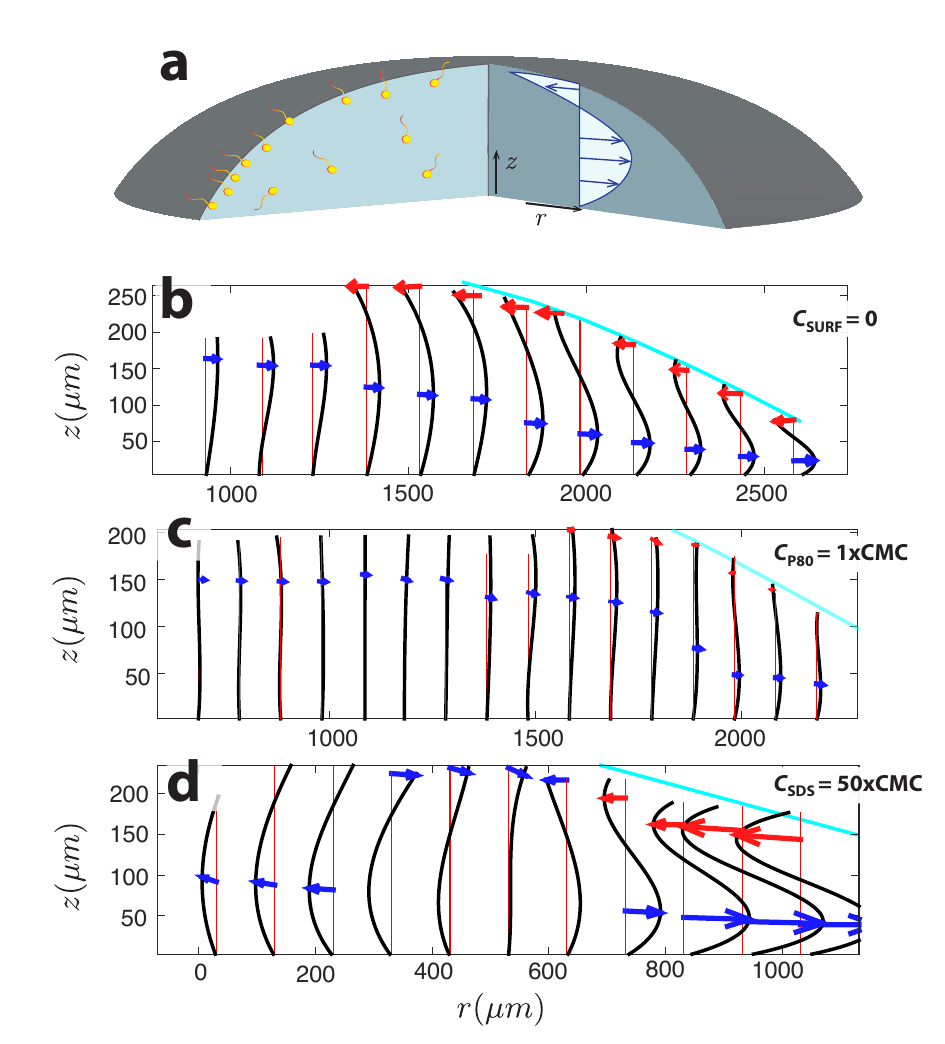}
\caption{(a) Sketch of a sessile evaporating water droplet on a glass substrate. The velocity profile drawn inside corresponds to that experienced when no surfactants are present in the solution. (b) Experimental velocity profiles for the case of a surfactant-free droplet. (c) Experimental velocity profiles for the case of a droplet at $C_\mathrm{P80}=C^\mathrm{P80}_\mathrm{CMC}$.(d) Experimental velocity profiles for the case of a droplet at $C_\mathrm{SDS}=50 C^\mathrm{SDS}_\mathrm{CMC}$.}
\label{fig1}
\end{figure}

The role of surface active impurities at the droplet's surface has been also a topic of debate. On the one hand, their presence has often been assumed to compensate the thermal Marangoni flow by generating a counter-gradient of surface tension. \cite{Hu:2005dv,ristenpart2007influence} On the other hand, recent studies show an enhancement of surface flow upon the addition of surfactants, \cite{Sempels:2013bi, still2012surfactant} even reporting coffee-stain effect reversal. However, it must be noted that since the visualization is normally performed through a projection in the image plane, particles at the surface and particles within the bulk look almost identical and therefore the conclusions of those reports must be carefully considered.
Such a lack of experimental data on the superficial flow is understandable given the complexity of the system. First, the droplet's surface is changing in time, decreasing its height linearly in time. Second, the droplet's surface is curved due to capillarity, which is pretty inconvenient for most of the visualization techniques, based on flat object planes with low depth of focus. Some efforts have been recently made either using fast confocal microscopy \cite{Bodiguel:2010cy} or using optical coherence tomography.\cite{trantum2013cross} Unfortunately, both lack a proper temporal resolution and are only able to track particles in short periods of time (optical coherence tomography) or scanning one plane at a time (confocal microscopy).

The aim of this Paper is to investigate how the evaporation-induced flow in evaporating sessile water drops is affected by the presence of surfactants. The experiments are performed by tracking the three-dimensional position of micro-particles dispersed in evaporating water droplets with unprecedented high spatial and temporal resolution. For the first time we are able to fully resolve the thermal Marangoni flow developed on surfactant-free water droplets, being able to calculate shear stresses and temperature differences that match the analytical results on the literature,\cite{Hu:2005dv} but contradicts the predicted temporal evolution. Finally, our results reveal radically different surface dynamics when soluble surfactants of different nature are introduced in the solution above their critical micellar concentration, which are consistent with recent studies on soap film formation. \cite{champougny:2015ij}

\paragraph{\bf Experimental Setup.- }

Experiments are performed with sessile evaporating water droplets in an open aluminium chamber at atmospheric conditions, with humidity between 40\%-45\% and 20$\SIC$ temperature. A droplet is gently deposited on a glass slide, which is at the same time held by a thick aluminum holder (at room temperature) that serves also as heat sink. \textcolor{black}{The glass slides are cleaned with ethanol and triply deionized water before each experiment. Water droplets on clean glass slides partially wet the substrate, with initial contact angles that may vary between 15$^\circ$ and 30$^\circ$}. The chamber, which is also in contact with the holder, protects the droplet from air currents that strongly disturb the droplet's surface flow. The temperature at the aluminium chamber and holder is monitored during the evaporation process to make sure that there are no significant external temperature variations. There is not additional control of the temperature in the system. 
The droplet contains a very low concentration of fluorescent polystyrene (PS) particles (below 0.001~\% w/w). \textcolor{black}{Such a low particle concentration is necessary for performing volumetric particle tracking. The polystyrene particles are supplied by $\mathrm{Microparticles~GmbH}$ and are coated with sulfate groups to avoid aggregates. Although their hydrophobic character is partially mitigated by the coating, they still show low adherence towards the glass slide, which is quite convenient when they are used as flow tracers.} They have a nominal diameter of 2~$\SImum$ and density of $\rho^{{}}_\mathrm{ps} = 1050~\:\SIkg\:\SIm^{-3}$. The particles are fabricated and labeled with a proprietary fluorescent dye ($\mathrm{PS-FluoRed}$) to be visualized with an inverted epifluorescence microscope. Simultaneously, a side view of the droplet is obtained through a glass window of the chamber in order to measure the droplet profile in time and therefore obtain the contact angle evolution in time and the evaporation rate. Additional data regarding the experimental set-up, evaporation rates, and other technical details can be found in the supplementary material.

In order to analyze the effect of surfactants of different characters, we use two commonly-used water-soluble surfactants: \textcolor{black}{polysorbate} 80 (P80) and \textcolor{black}{sodium dodecyl sulfate (SDS)}. P80 is a non-ionic surfactant often used as emulsifier and SDS is an anionic surfactant used as detergent in many products. Note that P80 is a much larger and complex molecule than SDS. The values of the critical micellar concentration (CMC) employed are $C_\mathrm{P80}^\mathrm{CMC}=0.012~\:\mathrm{mM}$ and $C_\mathrm{SDS}^\mathrm{CMC}=8.2~\:\mathrm{mM}$. We perform experiments studying the flow inside the droplet for surfactants concentrations below and above the CMC. For reasons that will be explained below, experiments with P80 are performed on standard glass slides with droplets of typical volumes of approximately 5~$\SImuL$, while typical volumes of approximately 1~$\SImuL$ are used with experiments on SDS, with the droplets gently deposited on teflon-printed glass slides with circular grooves of 2-mm diameter.

\paragraph{\bf 3D particle tracking.-} The particle trajectories and velocities are measured using astigmatism particle tracking velocimetry (APTV). \cite{Cierpka2011, rossi2014optimization} APTV is a single-camera particle-tracking method in which an astigmatic aberration is introduced in the optical system by means of a cylindrical lens placed in front of the camera sensor. Consequently, an image of a spherical particle obtained in such a system shows a characteristic elliptical shape unequivocally related to its depth-position $z$. Particle images are acquired using an inverted microscope Zeiss Axiovert in combination with a high-sensitivity sCMOS camera. A wide range of recording speeds from 0.01 fps up to 100 fps can be chosen. The experiments shown were recorded at 1 fps, which was enough to capture the particle motion at good temporal and spatial resolution. The optical arrangement consisted of EC Plan-Neofluar 10x/0.3 microscope objective lens and a cylindrical lens with focal length $f_\mathrm{cyl} = 300$~mm placed in front of the CCD sensor of the camera. Illumination is provided either by a pulsed diode-pumped laser or by a low-power continuous laser with 532~nm wavelength. This configuration provided a maximum measurement volume of about $1500\times 1500\times 300$~$\SImum^3$ with an estimated uncertainty in the particle position determination of $\pm$ 1~$\SImum$ in the $z$-direction and less than $\pm$ 0.1 $~\SImum$ in the $x$- and $y$-direction. More details about the experimental configuration and uncertainty estimation of the APTV system can be found in the supplementary information and in {Rossi et al.}\cite{rossi2014optimization}

\begin{figure*}
\centering
\includegraphics[width=.8\textwidth]{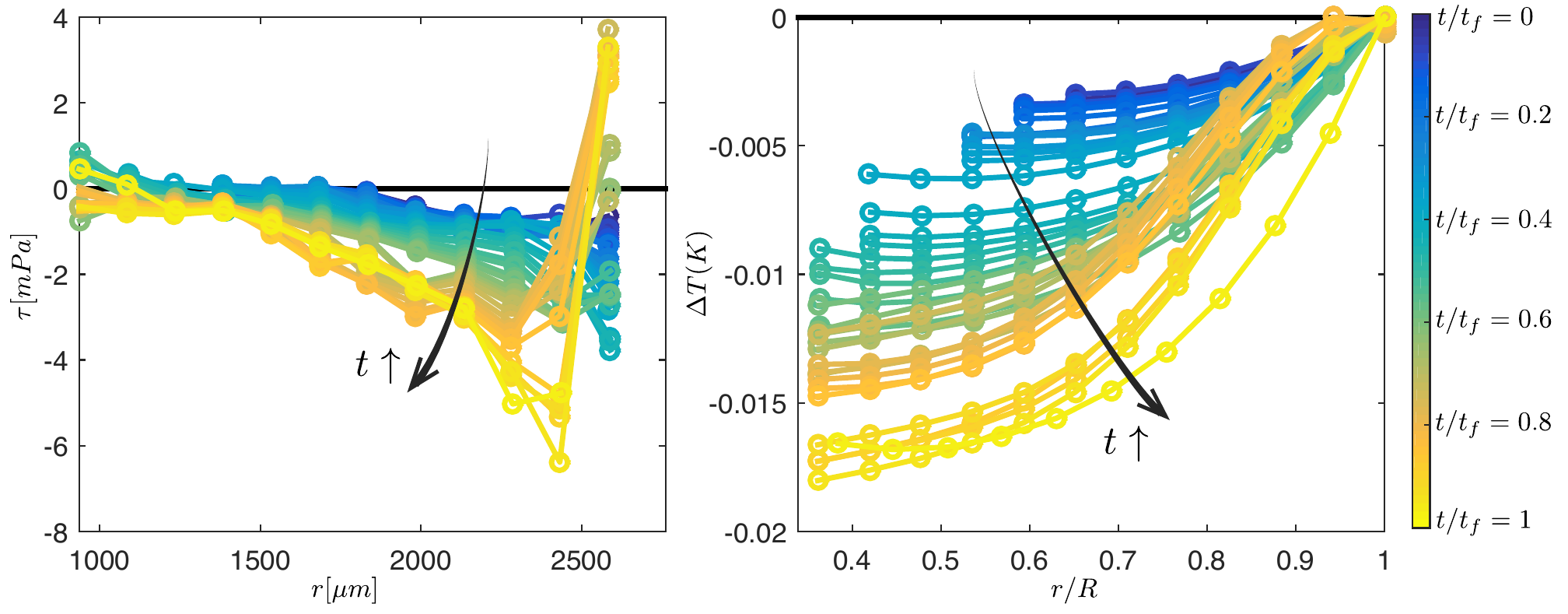}
\caption{Surfactant-free evaporating droplet: (a) Surface Shear stress $\mathrm{\tau}$ vs the distance to the center of the drop $\mathrm{r}$, as calculated from the velocity profiles in Fig. \ref{fig1}, at different time steps $t/t_f$. (b) Temperature difference, obtained by integration of the surface tension gradient field. Note that the reference $\Delta T=0$ has been set at $r/R=1$. Given the already low initial contact angles, the gradient along the droplet's surface can be represented with the radial coordinate $r$ with good accuracy.}
\label{fig2}
\end{figure*}

\paragraph{\bf Surfactant-free droplets} The flow within a sessile evaporating drop has been extensively investigated experimentally, numerically and analytically. \cite{Marin:2011hr, Deegan:1997vb, Hu:2005fq} However, the surface flow has remained unexplored experimentally mainly due to the difficulty of performing velocimetry close to a continuously-changing free surface. 
Using APTV, it is possible to accurately measure the thermally-induced Marangoni flow that develops spontaneously at the droplet's free surface. A typical velocity profile is depicted in Fig. \ref{fig1}b. Three important remarks need to be done about these results: (1) The surface Marangoni flow is directed towards the center of the droplet, with its maximum located close to the contact line, and decays to zero at the center of the drop. (2) Contrary to what has been predicted by models and simulations, \cite{Hu:2005dv, ristenpart2007influence} the Marangoni flow increases during the whole evaporating process (see video in the supplementary information). (3) It is well-known that the capillary-driven bulk flow scales linearly with the droplet radius.\cite{deegan2000contact} Interestingly, the same trend is observed for the surface Marangoni flow, which seem to scale linearly with the droplet radius. 
A typical velocity profile for surfactant-free droplets is plotted in Fig. \ref{fig1}b. The black thick lines correspond to polynomial fittings of the dimensionless radial component of the particle velocity $v_r(z)/v_o$ performed along rings separated by a radial distance $\delta r=0.1R$ from each other. The radial velocity values are normalized with $v_o=D\Delta{C}/{R\rho_l}$, where $D$ is the vapor diffusivity, $\Delta{C}=C_{\infty}-C_{S}$ is the vapor concentration difference, $R$ is the droplet radius and $\rho_{l}$ the liquid density. The choice of such scale comes from the fact that bulk velocity is directly proportional to the droplet's evaporation rate. Blue arrows depict the maximum of the bulk flow on each radial position, and red arrows the maximum value of the surface flow on each radial position. In order to choose the values of surface and bulk velocity at each radial position, an algorithm fits the velocity profile $v_r(z)$ to a third order polynomial and finds the local maxima/minima of the velocity profile close to the droplet's surface (if it is already within the measurement volume) and that closer to the substrate. \textcolor{black}{The largest source of error comes from the determination of the particles z-position ($\pm 1 \mathrm{\mu m}$) and from the particle's Brownian motion. In order to minimize such errors, only long particle trajectories are taken into account and the velocity profiles are obtained with thousands of particles. As a result $v_r(z)$ is given with an estimated margin of error of 15 \%, which results in 20\%-30\% error in the calculated velocity gradient values $\partial_z v_r(z)$ . }

\begin{figure*}[t]
\centering
\includegraphics[width=.8\textwidth]{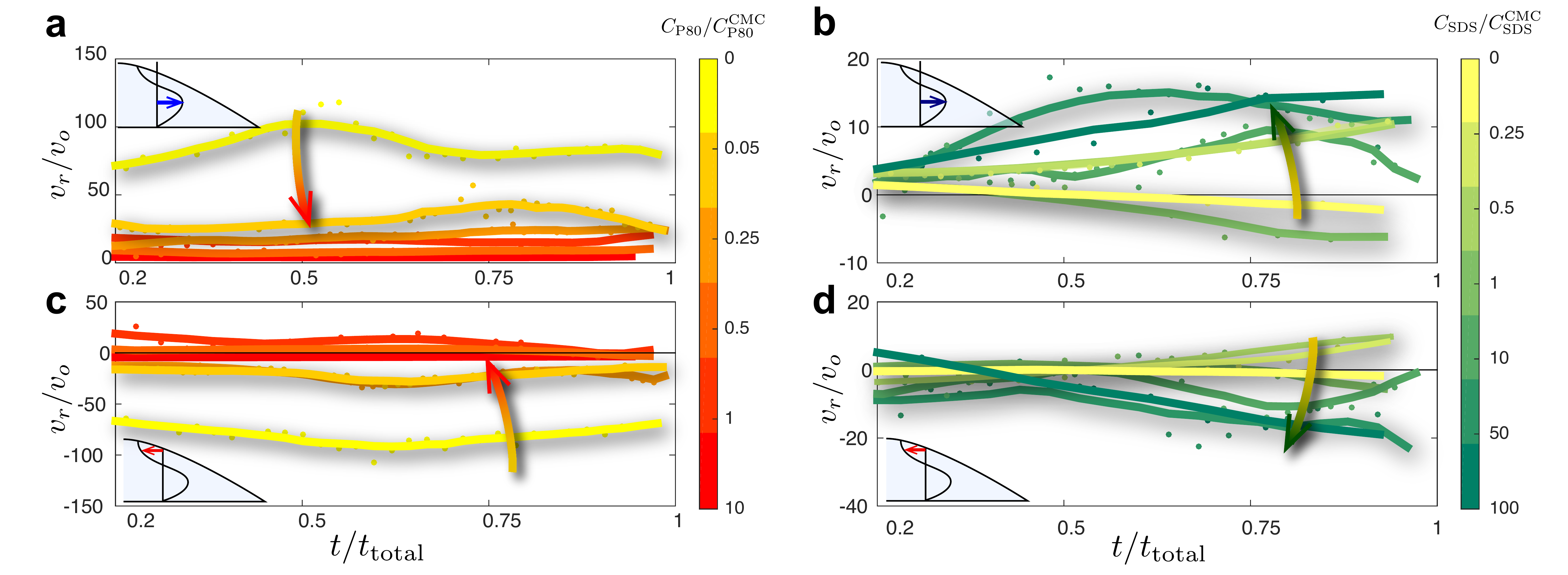}
\caption{Radial velocity vs. relative evaporation time for evaporating droplets droplets with Surfactants P80 (a, c) and SDS (b,  d). Bulk velocity vs. relative evaporation time are shown in bluish colors in (a) and (b): Lighter blue corresponds to a surfactant-free droplet, black corresponds to (a) $C_\mathrm{P80}=10C_\mathrm{P80}^\mathrm{CMC}$ and (b) $C_\mathrm{SDS}=50C_\mathrm{SDS}^\mathrm{CMC}$. Surface velocity vs. relative evaporation time are shown in reddish colors in (c) and (d): Lighter red corresponds to a surfactant-free droplet, black corresponds to (c) $C_\mathrm{P80}=10C_\mathrm{P80}^\mathrm{CMC}$ and (d) $C_\mathrm{SDS}=50C_\mathrm{SDS}^\mathrm{CMC}$. Note the totally opposite trends that flow velocity shows as the surfactant concentration of the different surfactants increases: P80 reduces both bulk and surface flow, while SDS increases it dramatically.}\label{fig3}
\end{figure*}

The main advantage of employing a three-dimensional tracking technique in such a system is the possibility of calculating the shear at the surface. In the case at hand, any stress that occurs at the surface is originated by a surface tension gradient, i.e. we can define the surface stress $\mathrm{\tau}$ as 

\begin{equation}
\tau=\mu {\partial_z v_r}|_{z=h(r)}=\partial_r \gamma.
\label{eq:tau-st}
\end{equation}

Which gives us a direct relationship between the experimentally measured velocity gradients ${\partial_z v_r}|_{z=h(r)}$ and the surface tension gradient $\partial_r \gamma$. \textcolor{black}{Note that in the following we will assume low contact angles and lubrication approximation such that the gradient along the surface can be calculated using $r$. This is indeed the case for most experiments, in which the contact angle drops below 10$^\circ$ typically at $t>0.25t_f$.} In Fig. \ref{fig2}(a) we show the measured surface shear stress as a function of the radial distance from the center at different times of the process: (1) Shear decreases as we approach to the center of the droplet, as expected by radial symmetry. (2) It reaches a maximum value close to the contact line, with an almost linear trend at early times, and non-linear at late times. (3) Surface shear stress is directed towards the center of the drop ($\tau<0$) at almost all times. Only at very late times ($t>0.9t_f$) and very close to the contact line ($|r-r_\mathrm{cl}|<100~ \:\mathrm{\mu m}$), a sudden change of sign of the shear occurs with significantly high and positive values. In order to interpret these values, we calculate the surface tension difference responsible for such thermal Marangoni stress by integrating Eq. \ref{eq:tau-st} in the available range of ${r}$. Furthermore, assuming that the source of the surface stress is purely thermal, we can also calculate the temperature difference by simply taking into account the chain rule
\begin{equation} 
\frac{d \gamma}{d r}=\frac{d \gamma}{dT} \frac{d T}{d r},
 \label{eq:chain}
\end{equation}
where $d\gamma/dT=-0.1657 ~\:\mathrm{mN/{m\cdot K}}$ has been taken from the literature.\cite{RubberBibble} By integrating equation \ref{eq:chain}, we can obtain values for the relative temperature difference: $\Delta T(r)=T(r)-T_{r=R}$, which is shown in Fig. \ref{fig2}(b). The temperature difference is obtained for convenience respect to its value at the contact line, as will be discussed further below. Two important comments need to be done regarding Fig. \ref{fig2}(b): On the one hand, the maximum temperature difference along the droplet surface is $\Delta T|_\mathrm{max}\approx$ 0.02~K, which fits quantitatively well with the maximum temperature gradient computed by Hu and Larson \cite{Hu:2005dv} in their numerical work and with the experiments and analytical models by Dunn et al. \cite{dunn2009strong}. On the other hand, all accepted models in the literature assume that the temperature difference along the surface should \emph{decrease} in time. At early times, the thermal influence of the substrate dominates, warming up the lower part of the droplet ($\Delta T>0$). As the droplet becomes thinner, the evaporative cooling increases, eventually inverting the temperature gradient ($\Delta T<0$). This should be accompanied by a reversal of the Marangoni flow at the surface, oriented towards the contact line below a critical contact angle. \cite{ristenpart2007influence} However, as we can see from Fig. \ref{fig2}, the experimental results show quite the opposite trend: the temperature difference $\Delta T$ between the contact line and the center of the drop \emph{increases} as the droplet evaporates, reaching its maximum temperature difference (and also maximum shear) in the last stages of evaporation. Note that our measurements do not permit to infer the absolute temperature at a certain point, but only relative temperatures differences can be obtained. Therefore, the temperature at the contact line $T_{r=R}$ might not be constant in time. Nonetheless, given the fact that the substrate is the only heat source in contact with the droplet, it is reasonable to choose the contact line as reference for measuring the temperature difference $\Delta T$. More detailed and direct measurements of the local variations of temperature in an evaporating droplets have been studied by Sefiane et al.\cite{sefiane2009}

The fate of those particles that do not get stuck at the contact line is particularly interesting to discuss since they follow the surface flow towards the center of the drop until the thermal Marangoni flow almost vanishes. Thermal Marangoni flows in evaporating flows are normally illustrated in the literature as a recirculating flow pattern. In such a way, one would expect that a particle dragged from the contact line by the Marangoni loop will be carried again towards the contact line by the bulk flow. Reality is more complex: the flow strength along such a recirculating pattern is not homogoneus, with higher velocities close to the contact line, and negligible velocities close to the droplet's center. Those particles dragged along the surface by the Marangoni flow towards the droplet's center get ``trapped'' at the surface or most likely close to it, where both surface flow and bulk flow are almost negligible. This is indeed the origin of the so-called ``skin'' that Deegan \cite{deegan2000pattern} and later others described and studied in the context of polymer solutions, \cite{Manukyan:2013gh} and that remains until the last instants of the process. In low concentration suspensions as it is our case, particles do not form a film or a skin but simply remain close to the surface, ultimately approaching the substrate at the same pace as the surface does. \cite{trantum2013cross} In the last instants, when the capillary flow increases dramatically, those particles will most likely be carried away towards the contact line. \cite{Marin:2011ts} However, when the concentration of particles is high enough and/or the particles interact forming clusters, they might end up in the central region of the droplet when the solvent is completely evaporated. This is a clear example of the importance on the particle and substrate physicochemistry \cite{moraila2012effect,moraila2013role,NogueraMarin:2015bu} when analyzing the particle distribution in dried deposits, which is often ignored giving a dominant role to evaporation-induced flows. 

\paragraph{\bf Droplets with P80.- }
Figure \ref{fig1}b shows how the flow changes dramatically when P80 is added at its CMC: The velocity profiles become almost flat, i.e. both bulk flow and surface flow are substantially reduced. This is more evident when looking at Fig. \ref{fig3}, where the average bulk and maximum surface velocities have been plotted in time. It clearly shows that droplets with concentration of P80 above $C_\mathrm{P80}^\mathrm{CMC}$ present a totally ``rigid'' surface (no surface mobility). Such a rigid surface creates a non-slip condition at the droplet's surface, increases the viscous dissipation and substantially reduces the bulk velocity. For this set of experiments larger droplets have been chosen ($R \approx$ 2.5 - 3~mm). Since the evaporation rate scales linearly with the droplet radius, \cite{Gelderblom:2012by} it is also observed that larger droplets yield larger values of the Marangoni flow. Therefore, in order to have a more notorious decrease of surface and bulk motion, larger droplets have been chosen for this set of experiments. Note also that the velocity reduction cannot be due to an increase of viscosity in the droplet since the CMC in the surfactant P80 is achieved at very low concentrations (0.012~mM). 

At even higher concentration values of P80 (larger than 100 CMC) a more complex behavior has been observed where the flow actually inverts its direction: the surface flow is directed outwards and the bulk flow inwards. This is exactly the opposite behavior as reported by Sempels et al. \cite{Sempels:2013bi} with standard video microscopy footage using the same surfactant. The most likely explanation is a misinterpretation of their particles' z-position due to the lack of 3D information. We noted that at such concentrations also deposits of precipitated surfactant are observed at the contact lines, leading to gelation and deforming the droplet's shape. The processes in those cases of extreme concentrations and gel-like deposits are even more complicated to study and interpret \cite{poulard2007control,okuzono2009final,Manukyan:2013gh} and will not be considered.

\begin{figure*}
\centering
\includegraphics[width=.8\textwidth]{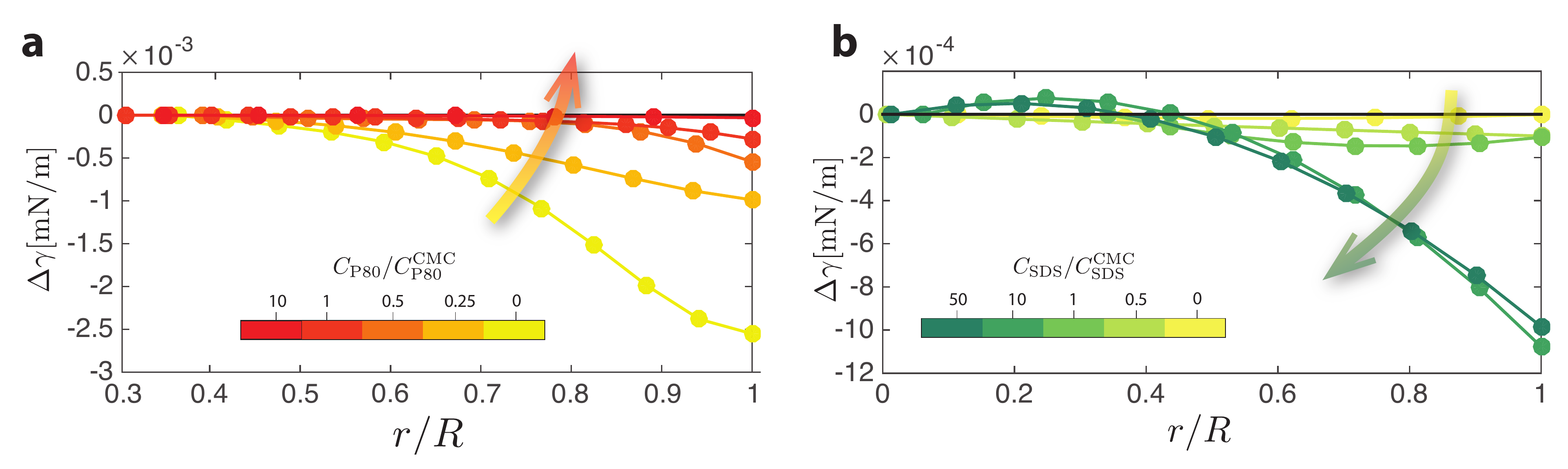}
\caption{ Surface tension differences along the droplet's radial coordinate upon addition of surfactants, obtained by integration of the surface shear stress field. (a) Results on large droplets containing P80. At increasing concentrations the shear stresses vanish and the surface becomes immobile. (b) Results on smaller droplets containing SDS. Note that smaller droplets show a substantially reduced flow when clean, but a timid enhancement of the flow is observed up to the critical micellar concentration of SDS, while a sudden change occurs for concentrations above it and up to $C_\mathrm{SDS}=50 C^\mathrm{CMC}_\mathrm{SDS}$.}\label{fig4}
\end{figure*}

\paragraph{\bf Droplets with SDS.- }

The addition of ionic surfactant SDS below the CMC has little effect on the observed flow, but a clear transition is observed above the CMC. A typical velocity profile of a droplet saturated of SDS is shown in Fig. \ref{fig1}d: Both the surface flow and the bulk flow are significantly enhanced close to the contact line, with similar characteristics as in the case of surfactant-free droplets. The main difference with surfactant-free droplets is that the radial flow inverts in an area close to the droplet's center, generating an internal recirculating pattern with opposite vorticity as the ``external'' one (also visible in Fig. \ref{fig1}d). Interestingly enough, this recirculation patterns do have a very homogeneous strength and therefore, particles are seen to recirculate back and forth in these loops (in contrast with those generated by thermal gradients).  
The value of the bulk velocities found is approximately ten times larger than the case without surfactant, and twenty times larger for the surface velocity flow. Such a behavior is only observed for concentrations spanning from 1 to 100 CMC. Similar behavior has previously been described by other authors, \cite{Sempels:2013bi, still2012surfactant} although the data given was based on the apparent size of the vortices or on projected motion of bacteria.
Note that droplets with smaller radius ($R\approx$~1~mm) have been used in this set of experiments. It is worth mentioning that in the absence of surfactant, the surface thermal Marangoni flow is to weak to be discerned from the particle Brownian motion in such small droplets. Only by observing the coherent motion of the particles at longer time scales it is possible to perform measurements and quantify such a flow. Probably for this reason, some authors have often reported the absence of thermal Marangoni flows in evaporating capillary water droplets at room conditions. As expected, the addition of P80 in such small droplets only makes the system even less dynamic.

\paragraph{\bf Discussion.-}

By integrating the surface shear stress in the case of surfactant-laden droplets, we can compute the surface tension difference. This is done for the different surfactant concentrations explored, and for one single time interval close to the end of the evaporation process, when the motion inside the droplet is the highest. Figure \ref{fig4} shows the measured surface tension differences, relative to the center of the drop $r/R=0$. 

First, the addition of surfactant P80 on large droplets tends to reduce the shear caused by the temperature gradients. When the concentration reaches the CMC, the surface shear is hardly measurable. On the other hand, the addition of SDS in small droplets tends to increase the surface shear moderately below the CMC (Fig. \ref{fig4}), but a transition clearly occurs above circa 2 CMC: surface tension drops dramatically at the contact line, increasing the motion in its vicinity, but at the same time surface tension seems to increase slightly with respect to the drop's center at $r/R\approx0.2$, creating an internal counter-rotating loop. 
Note that although the maximum surface tension gradient is extremely low (1 $\mathrm{\mu N/m}$ per mm), they are able to generate a reproducible surface flow in the range of 10 $\mathrm{\mu m/s}$. This value is consistent with experiments on film formation \cite{champougny:2014vp} or by forcing surface tension gradients, \cite{Roche:2014dh} in which the typical velocity scale found is in the order of 1 $\mathrm{mm/s}$ for surface tension differences of 1 mN/m.

The results shown are unprecedented and raise a number of questions. The first one being: Why is the behavior among different surfactants so remarkably different? This must necessarily be related with the different nature of the two surfactants: P80 is a large and non-ionic surfactant, and its surface pressure reaches equilibrium typically within the time of evaporation (15-20 minutes) at the CMC. \cite{maldonado2007surface} Research on the adsorption of soluble non-ionic surfactants at interfaces \cite{eastoe1997dynamic} shows that the surface tension decay follows a mixed diffusion-activation adsorption mechanism. Early times are typically dominated by a faster diffusion, specially on those surfactants as P80 with a low CMC value. Combined with the fact that surfactant monolayers of P80 have a relatively high compaction at the CMC and almost negligible elastic effects,\cite{Jaishankar:2011iua} we can conclude that P80 forms a stable and rigid monolayer in the early instants of evaporation, therefore reducing the shear stress on the surface and the flow motion within the droplet. Such surfactant-induced increase of surface rigidity is also responsible for the enhancement in soap film formation. \cite{champougny:2014vp} It should be noted that, due to the cylindrical symmetry of the flow and the system itself, the surface is actually being more compressed than sheared, and therefore the most relevant surface microrheological variables are rather dilatational than shear viscosity and elasticity. \cite{Zell:2014dz}

On the other hand, SDS is known to break the rigidity of surfaces stabilized by proteins and enhance foam drainage. \cite{Gauchet:2015dm,saint2004quantitative} Even when the surface is covered by surfactant, it is able to remain mobile \cite{bonfillon1993viscoelasticity} and consequently concentration gradients can easily be generated. Such concentration gradients will become larger as the surfactant concentration increases. 
Regarding the direction of the gradient, SDS must preferentially be adsorbed at the contact line due to both the higher surface-to-volume ratio and the evaporation-driven flow, which generates the surface flow observed towards the center of the drop. Such a flow tries to equilibrate the surface tension gradient along the surface, but it is clearly slower than the replenishment of surfactant due to the bulk's convective flow. However, the whole surface must be quickly almost completely covered by surfactant, since the surface tension differences are in the order of 1~$\mathrm{\mu N/m}$. 

It is important to point out that even though Marangoni flows as those observed in our experiments have comparable strength as the capillary-driven flow responsible for the coffee-stain effect, they are not able to reverse the particle deposits \cite{Hu:2006tx,ristenpart2007influence}: During the evaporation of the droplet, Marangoni flows lead the particles to the central part of the droplet, where they are still sensitive to the fluid motion. At the end of the process, most particles dispersed in the liquid will always be dragged towards the contact line in the last instants of evaporation due to the so-called ``rush-hour effect'' (i.e. due to mass conservation)(See videos in supplementary files). As a matter of fact, little correlation between the Marangoni flow and the deposition patterns is found in our experiments. Larger correlations have recently been described on other factors not explored in this study as the wettability of the substrate, the particle-substrate adherence, or the particle electrical charge.\cite{moraila2012effect,moraila2013controlling,moraila2013role,NogueraMarin:2015bu} The most efficient way to avoid coffee-ring patterns in an evaporating drop is by making the contact line mobile, and therefore eliminate its self-pinning. This occurs naturally on hydrophobic substrates and using particles with low adhesion. By doing that, the droplet's contact line recedes as it evaporates, dragging particles along until they concentrate in the center of the droplet. Note that this mechanism occurs in the last milliseconds of evaporation and it is completely independent of the flow within the droplet. Another way to do this in an active way uses electrowetting to mobilize the contact line. \cite{eral2011suppressing,eral2012say}


\paragraph{\bf Conclusions and Outlook.-}

By using a defocusing particle image technique (APTV), we have been able to measure the flow inside evaporating water droplets and close to the droplet's surface in the presence of surfactants with unprecedented spatial and time resolution. Such a technique allows us to measure not only the flow inside the droplet, but also the surface shear that develops at the surface, and consequently, the surface tension difference. Our results with surfactant-free droplets clearly show the presence of a thermal Marangoni flow, whose strength is consistent with the values predicted by models and simulations. \cite{Hu:2005dv} Interestingly, the thermal Marangoni flow measured increases as the droplet becomes thinner and thinner, which strongly disagrees with theoretical models and simulations from the literature.  
 
On the one hand, the addition of the non-ionic surfactant P80 tends to homogenize those surface tension gradients caused by the temperature gradient. Above the CMC, the surface tension gradients are almost completely vanished and the surface becomes totally immobile. This effect  also reduces the bulk flow strength, in a similar way as certain surfactants slow down the drainage of soap films (or enhance the film extraction) due to the surface rigidity. On the other hand, the addition of the ionic SDS in small droplets tends to increase the surface tension moderately below the CMC (Fig. \ref{fig4}), but a transition clearly occurs above 2 {CMC}: surface tension drops at the contact line, increasing the motion close to the contact line up to values that are actually comparable to the thermal Marangoni flow in larger droplets. The behavior of SDS in evaporating droplets is qualitatively comparable to the enhancement reported in film drainage experiments, in which SDS always seems to yield a plug flow, with a totally mobile surface even at the highest bulk concentrations. 

In conclusion, our experiments give further evidence of the complexity hidden in such an apparently simple and common system as an evaporating sessile water droplet. The results also evidence the limitations of many of the models and assumptions that have been made in the past, and open a door to new models and simulations, as well as encourages to different and more appropriate experimental approach.

\paragraph{Acknowledgments.-}
Author Contributions: {AM developed the concept of the study. AM and RL performed the experiments. AM, RL and MR processed the data and conducted the analysis. AM interpreted the data and wrote the manuscript. CJK contributed to the development of the technique and interpreted the  data.} The authors acknowledge financial support by the Deutsche Forschungsgemeinschaft grant KA 1808/12-1. The authors would like to acknowledge fruitful discussions with Rune Barnkob.



\bibliographystyle{unsrtnat}

\end{document}